\documentclass[conference]{IEEEtran}

\ifCLASSOPTIONcompsoc
\usepackage[compress]{cite}
\else
  \usepackage{cite}
\fi
\ifCLASSINFOpdf
\else
\fi
\usepackage{pgfplots}
\pgfplotsset{compat=1.17}

\usepackage{pgfplots} 
\usetikzlibrary{patterns}
\usepackage{booktabs} 
\usepackage{url}
\usepackage{xspace}
\usepackage[font=normal]{caption}
\usepackage{subcaption}
\usepackage{float}
\usepackage{textcomp}
\usepackage{multirow,makecell}
\usepackage{pifont}
\usepackage{tikz}
\usepackage{bbm}
\usepackage[inline]{enumitem}
\usepackage{adjustbox}
\newcommand{\etal}{{\em et al.}\xspace}

\newcommand{\BfPara}[1]{{\noindent\bf#1.}\xspace}

\usepackage{xcolor}

\usepackage{listings}
\usepackage{tikz}
\usetikzlibrary{tikzmark}
\usetikzmarklibrary{listings}
\usepackage{amsmath}
\usepackage{amssymb}

\usepackage{amsmath}
\usepackage{graphicx}
\usepackage{makecell}
\usepackage{algorithm}
\usepackage{algorithmic}

\usepackage{colortbl}
\definecolor{darkgreen}{rgb}{0.0, 0.3, 0.13}
\definecolor{darkred}{rgb}{0.2, 0.0, 0.13}

\usepackage[table]{xcolor}

\definecolor{best}{RGB}{198,239,206}
\definecolor{second}{RGB}{255,235,156}
\definecolor{worst}{RGB}{255,199,206}

\newcommand{\best}[1]{\cellcolor{best}\textbf{#1}}
\newcommand{\second}[1]{\cellcolor{second}#1}
\newcommand{\worst}[1]{\cellcolor{worst}#1}

\usepackage[most]{tcolorbox}
\tcbset{textmarker/.style={%
        parbox=false,boxrule=0mm,boxsep=0mm,arc=0mm,
        outer arc=0mm,left=3mm,right=3mm,top=3pt,bottom=3pt,
        toptitle=1mm,bottomtitle=1mm}
        }
        
\newtcolorbox{blueBox}{textmarker,
    colback=blue!10!white}

\usepackage[english]{babel}
\usepackage{blindtext}
\usepackage{xspace}
\usepackage{filecontents}
\usepackage{multicol}

\lstdefinestyle{myStyle}{
  belowcaptionskip=1\baselineskip,
  breaklines=true,
  language=C++,
  showstringspaces=false,
  basicstyle=\footnotesize\ttfamily,
  keywordstyle=\bfseries\color{green!40!black},
  commentstyle=\itshape\color{purple!40!black},
  identifierstyle=\color{blue},
  stringstyle=\color{orange},
  numbers=left,
  firstnumber=1,
}

\usepackage{hyperref}
\hypersetup{
    colorlinks=true,
    linkcolor=blue,
    citecolor=green,
    urlcolor=blue,
    filecolor=blue,
    bookmarks=true,
    bookmarksopen=true,
    breaklinks=true,
    pdfauthor={author},
    pdftitle={title},
    pdfsubject={},
    pdfkeywords={}
}
\usepackage{tikz}
\usepackage{tabularx}
\usepackage{colortbl}
\usepackage{orcidlink}

\usepackage{pgfplots}
\pgfplotsset{compat=1.18}
\usepgfplotslibrary{groupplots}

\tcbset{
  condensedtakeaway/.style={
    colback=blue!5,
    colframe=blue!75!black,
    fonttitle=\bfseries,
    title=Takeaway,
    boxrule=0.8pt,
    arc=2mm,
    left=2mm,
    right=2mm,
    top=1mm,
    bottom=1mm,
    enhanced,
    sharp corners=south,
  }
}

\begin{document}

\title{\LARGE Concept Drift Adaptation Using Self-Supervised and Reinforcement Learning In Android Malware Detection}

\author{Ahmed Sabbah$^\dagger$, Mohammed Kharma$^\dagger$, Mohammad Alkhanafseh$^\dagger$, Samer Zain$^\dagger$, Radi Jarrar$^\dagger$, David Mohaisen$^\ddagger$\\
$^\dagger$Birzeit University \hspace{10mm} $^\ddagger$University of Central Florida}

\markboth{}%
{Sabbah \MakeLowercase{\textit{et al.}}: Concept Drift Adaptation Using Self-Supervised and Reinforcement Learning In Android Malware Detection}

\IEEEtitleabstractindextext{
\begin{abstract}
Android malware detectors often degrade after deployment because of concept drift, while full retraining at each maintenance step is costly. We propose a chronological adaptive maintenance framework that models deployment-time maintenance as a sequential decision problem. The framework learns a stable latent representation through self-supervised learning during initialization, freezes the encoder, measures latent drift in the fixed representation space, and performs lightweight downstream adaptation using a trainable adapter and classification head. A proximal policy optimization controller selects low-cost maintenance actions based on the detector state, including current utility, retention on a fixed memory set, latent drift indicators, and update cost. We evaluate the framework under a causal deployment-style protocol on emulator and real Android malware datasets with static and dynamic features. Results show that the RL controller provides a strong cost-aware adaptation strategy, consistently remaining among the top-performing policies while achieving a favorable balance between temporal performance, memory retention, and maintenance cost under non-stationary deployment conditions.
\end{abstract}

}

\maketitle

\IEEEdisplaynontitleabstractindextext

\IEEEpeerreviewmaketitle

\section{Introduction}\label{sec:introducion}

Android is the dominant mobile operating system worldwide, accounting for 67.46\% of the global market in March 2026~\cite{StatCounterMobileOS2026}. Its scale and prevalence make it a persistent target for mobile threats. According to Kaspersky, attacks on Android smartphone users in the first half of 2025 were 29\% higher than in the first half of 2024 and 48\% higher than in the second half of 2024~\cite{KasperskyPressH12025}. As a result, machine learning has become a widely adopted approach for Android malware detection using static, dynamic, or hybrid features. However, the effectiveness of these detectors depends not only on their initial accuracy but also on their ability to remain reliable as the Android ecosystem evolves. In supervised learning, concept drift refers to changes in the relationship between input variables and target labels over time~\cite{GamaZBPB14}. In Android malware detection, this challenge is particularly severe because malware behavior, benign application behavior, platform restrictions, API usage, and feature distributions continuously evolve~\cite{AbusnainaASAJSM25}. Consequently, detectors trained on historical data may perform well during development but degrade substantially after deployment.

A growing body of work has shown that this degradation is persistent and driven by multiple factors. Early studies highlighted the challenge of concept drift in Android malware detection and explored ensembles and online adaptation to sustain performance over time~\cite{HuMZLYL17,XuLDCX19}. Later work examined the causes of temporal degradation more closely, showing that timestamping choices, cross-device behavior, malware family evolution, benign application changes, and the distinction between feature-space and data-space drift all affect long-term detector reliability~\cite{Guerra-ManzanaresB22a,Guerra-ManzanaresLB22,Guerra-ManzanaresLB22b,ChenZKYCPPCW23,ChowKLCAP23}. More recent studies reinforced these findings through large longitudinal benchmarks, showing that concept drift remains widespread across datasets, feature types, and detector families~\cite{HaqueHKAATR25,SabbahJZM25}. Collectively, these results suggest that Android malware detection should be treated not as a one-time training problem, but as a long-term maintenance problem under temporal distribution shift.

To address this issue, prior work has explored several adaptation strategies. Ensemble and online learning methods incrementally update detectors as new data arrive~\cite{HuMZLYL17,XuLDCX19,KanPPC21}. Transfer learning approaches reuse prior knowledge to improve adaptation to newer samples~\cite{FuDG21,GarciaDC23}. Active learning and pseudo-labeling methods reduce annotation costs while supporting continuous retraining~\cite{ChenDW23,MolinaCoronadoMMM23,AlamFMR24,MuzaffarHZL26,AlamPR2025}. Rejection-based and optimization-oriented methods identify uncertain or drifted samples, improve training, and enhance robustness under evolving data distributions~\cite{PendleburyPJKC19,BarberoPPC22,MailletM23,HeLQRC25}. More recently, self-supervised learning (SSL) and reinforcement learning (RL) have also emerged in this space. MADCAT showed that self-supervised masked autoencoding with test-time adaptation can improve robustness under concept drift in Android malware detection~\cite{RohKKVH2025}, while DRMD demonstrated that reinforcement learning can support drift-aware decision making in time-aware malware detection settings~\cite{McFaddenFDHMPP25}.

\noindent{\bf Existing approaches do not jointly address adaptation, retention, and cost-aware maintenance under temporal drift.} Despite this progress, an important gap remains. Most prior studies either analyze the causes of drift or adapt detectors through repeated retraining, continual relabeling, or full-model updates. Few approaches jointly address adaptation and retention within a deployment-oriented framework, and recent SSL and RL methods are typically applied in isolation. This creates a need for a unified chronological maintenance framework that can adapt to new drifts, preserve previously learned behaviors, and control update costs. In this study, we address this gap by proposing a chronological adaptive maintenance framework for Android malware detection under a temporal distribution shift. The framework is designed to reflect a deployment-style setting in which a detector is initialized from early time windows and then maintained as later windows arrive. Instead of retraining the entire model repeatedly, we learn a stable latent representation during initialization using SSL. An RL controller then selects cost-aware maintenance actions based on drift signals, current detection behavior, retention performance, and action history. It also adopts a realistic evaluation protocol by separating adaptation and evaluation within each deployment window and measuring not only the current window performance, but also retention on a fixed memory set and the cumulative cost of maintenance actions.

\BfPara{Contributions} This work makes several contributions. First, we formulate Android malware detection under temporal distribution shift as a \emph{chronological adaptive maintenance} problem, where detectors are maintained over deployment windows instead of being repeatedly retrained from scratch. Second, we propose a unified framework combining self-supervised learning and reinforcement learning: SSL learns a stable latent representation during initialization, after which deployment-time updates are limited to a lightweight adapter and classification head. Third, we introduce latent drift monitoring using previous-window and initialization-reference drift in the frozen representation space. Fourth, we develop a maintenance policy that selects among multiple cost-aware actions, including keep, head-tune, adapter-tune, joint-tune, and reset-adapter. Finally, we adopt a deployment-oriented evaluation protocol that jointly measures adaptation, retention, and maintenance cost under chronological evaluation.

\section{Related Work}
\label{sec:related}

Concept drift is a well-established challenge in supervised learning and refers to temporal changes in the relationship between features and labels~\cite{GamaZBPB14,ChenZKYCPPCW23,ChowKLCAP23,FuDG21,GarciaDC23,HaqueHKAATR25,SabbahJZM25,Guerra-ManzanaresLB22,Guerra-ManzanaresLB22b,KanPPC21,MolinaCoronadoMMM23}. In Android malware detection, the problem is particularly severe because malware behavior, benign application behavior, platform APIs, and data collection conditions continuously evolve over time. Early studies showed that detectors trained on historical data degrade under chronological evaluation and therefore require explicit drift-aware maintenance rather than one-time training~\cite{HuMZLYL17,XuLDCX19,FuDG21,PendleburyPJKC19,BarberoPPC22,ChenDW23,MailletM23,RohKKVH2025,McFaddenFDHMPP25}. In the following, we review the main directions of prior work and position our framework relative to them.

\BfPara{Drift characterization and empirical evidence} A first line of work studies why Android malware detectors degrade over time. Prior studies showed that timestamping choices, feature-space versus data-space drift, malware family evolution, benign application changes, feature type, and collection settings all affect chronological robustness~\cite{Guerra-ManzanaresB22a,ChenZKYCPPCW23,ChowKLCAP23,Guerra-ManzanaresLB22,Guerra-ManzanaresLB22b}. More recent longitudinal benchmarks and empirical studies, including LAMDA~\cite{HaqueHKAATR25} and the work of Sabbah \etal~\cite{SabbahJZM25}, further showed that concept drift persists across datasets, models, and feature representations and must therefore be treated as a deployment-time problem, and not only at design time.

\BfPara{Adaptation through model updating}
A second line of work addresses drift through detector updates as new data arrive. Hu \etal proposed NBCS, which combines feature selection, sliding windows, and multiple sub-classifiers to sustain performance over time~\cite{HuMZLYL17}. DroidEvolver and DroidEvolver++ extended this direction with model pools, weighted voting, pseudo-label updates, and mechanisms for retiring outdated models~\cite{XuLDCX19,KanPPC21}. Transfer learning approaches reuse prior knowledge to adapt to emerging malware samples. Fu \etal fine-tuned an LSTM detector using augmented malware data, while Garcia \etal studied transfer learning under class imbalance across conventional classifiers~\cite{FuDG21,GarciaDC23}. Other work explored anomaly-detection-based online adaptation for settings with limited malicious labels~\cite{GarciaD23} and continual-learning regularization to reduce regression after updates~\cite{GhianiAPSMGPRB26}. Although these approaches delay degradation, most still depend on repeated retraining, additional labels, or broad parameter updates that may be costly in long-term deployment.

\BfPara{Label-efficient and optimization-oriented maintenance} Because continuous relabeling is expensive, several studies aim to reduce annotation cost while maintaining adaptation quality. Prior work explored active learning, pseudo-labeling, selective querying, and streaming-style retraining to support label-efficient adaptation under drift~\cite{ChenDW23,AlamFMR24,MuzaffarHZL26,AlamPR2025}. Other studies examined periodic versus drift-triggered retraining~\cite{MolinaCoronadoMMM23}, conformal prediction for rejecting uncertain or out-of-distribution samples~\cite{PendleburyPJKC19,BarberoPPC22}, and drift-resilient feature representations and training objectives~\cite{RochaDCDJ23,MailletM23,HeLQRC25}. While these approaches improve specific aspects of robustness, they do not formulate maintenance as a sequential decision problem that jointly considers adaptation benefit, retention, and update cost.

\BfPara{Closest directions to our framework} Most closely related to ours are self-supervised adaptation and reinforcement learning. MADCAT used self-supervised masked autoencoding with test-time adaptation to improve robustness under Android malware drift~\cite{RohKKVH2025}, while DRMD formulated time-aware malware detection as a deep reinforcement learning problem for drift-aware classification and rejection~\cite{McFaddenFDHMPP25}. However, MADCAT focuses on representation adaptation and DRMD focuses on decision optimization. Neither jointly addresses stable representation, lightweight maintenance, memory retention, and cumulative update cost within a unified framework.

\BfPara{Our work}
Prior work established that Android malware concept drift is persistent, multi-causal, and difficult to address through static training alone. Existing approaches have contributed important advances through drift analysis, online learning, transfer learning, active learning, rejection mechanisms, and recent SSL- or RL-based methods~\cite{HuMZLYL17,XuLDCX19,FuDG21,PendleburyPJKC19,BarberoPPC22,ChenDW23,MailletM23,RohKKVH2025,McFaddenFDHMPP25}. However, most methods still depend on frequent relabeling, full-model retraining, or objectives focused only on current adaptation performance. Our work instead targets chronological maintenance under deployment constraints by learning a stable latent space during initialization, restricting deployment-time updates to lightweight components, and using RL to select cost-aware actions based on drift, current performance, and retention.
\section{Background}
\label{sec:Background}

\subsection{Reinforcement Learning}
Reinforcement learning (RL) addresses sequential decision-making problems in which an agent interacts with an environment to maximize cumulative rewards over time~\cite{SuttonR18}. RL has been successfully applied in domains such as robotics~\cite{HoaCDTNDT20}, gaming~\cite{MnihKSRVBGRFOPB15,VuSEBD21}, and increasingly in cybersecurity applications, including malicious botnet detection in IoT environments~\cite{BakhshadPAWAT22}. RL problems are commonly formulated as Markov Decision Processes (MDPs), represented as:
\begin{equation}
\label{MDPquintuple}
MDP = (S, A, T, R, \gamma),
\end{equation}
where $S$ denotes the set of states, $A$ is the action space, $T$ defines the state transition probability distribution, $R$ is the reward function, and $\gamma \in [0,1]$ is the discount factor controlling the trade-off between immediate and future rewards.

The agent follows a policy $\pi$ that maps states to actions:
\begin{equation}
\pi(a|s) = P(a_t = a \mid s_t = s).
\end{equation}

Under policy $\pi$, the state-value function is defined as~\cite{SuttonR18}:
\begin{equation}
v^{\pi}(s) = \mathbb{E}_{\pi} \left[ \sum_{t=1}^{\infty} \gamma^{t-1} R_t \mid S_1 = s \right],
\end{equation}
while the state-action value function is:
\begin{equation}
q^{\pi}(s,a) = \mathbb{E}_{\pi} \left[ \sum_{t=1}^{\infty} \gamma^{t-1} R_t \mid S_1 = s, A_1 = a \right].
\end{equation}

Using the Bellman equations~\cite{SuttonR18}, the recursive value formulation becomes:
\begin{equation}
v^{\pi}(s) = \mathbb{E}_{\pi} \left[ R_{t+1} + \gamma v^{\pi}(S_{t+1}) \mid S_t = s \right].
\end{equation}

RL methods are categorized into model-based and model-free approaches where the first rely on explicit transition dynamics and the latter methods learn directly through interaction with the environment. Because cybersecurity environments are dynamic and evolve over time, model-free approaches are often more suitable. In this work, we use proximal policy optimization (PPO), a model-free RL algorithm.

\subsection{Proximal Policy Optimization (PPO)}
PPO is a policy-based reinforcement learning algorithm designed to achieve stable policy updates while maintaining sample efficiency~\cite{SchulmanWDRK17}. Unlike value-based methods such as Q-learning, PPO directly optimizes the policy through gradient ascent on the expected cumulative reward. PPO uses a clipped surrogate objective:
{\small
\begin{equation}
L^{CLIP}(\theta) =
\mathbb{E}_t \left[
\min \left(
r_t(\theta)\hat{A}_t,
\text{clip}(r_t(\theta),1-\epsilon,1+\epsilon)\hat{A}_t
\right)
\right],
\end{equation}
}
where
$r_t(\theta)=\frac{\pi_{\theta}(a_t|s_t)}{\pi_{\theta_{old}}(a_t|s_t)}$
is the policy probability ratio, $\hat{A}_t$ is the advantage estimate at $t$, and $\epsilon$ controls the clipping range.

PPO performs multiple mini-batch updates from sampled experiences while constraining harmful policy shifts through the clipped objective, combining the stability of trust-region methods with the simplicity of policy-gradient optimization.

\subsection{Concept Drift}
Concept drift refers to changes in the statistical properties of data over time in non-stationary environments~\cite{SchlimmerG86}. Let $P_t(X,Y)$ denote the joint distribution of features $X$ and labels $Y$ at time $t$. Concept drift occurs when:
\[
P_{t_1}(X,Y) \neq P_{t_2}(X,Y), \quad t_1 \neq t_2.
\]

Prior work categorizes concept drift into several forms, including sudden, gradual, and recurring drift~\cite{XiangZCW23}. In Android malware detection, concept drift arises because malware behavior, benign application behavior, API usage, and platform restrictions evolve over time, causing detectors trained on historical data to degrade after deployment.

\section{Methodology}
\label{sec:methodology}

\begin{figure*}[t]
    \centering
    \includegraphics[width=0.85\textwidth]{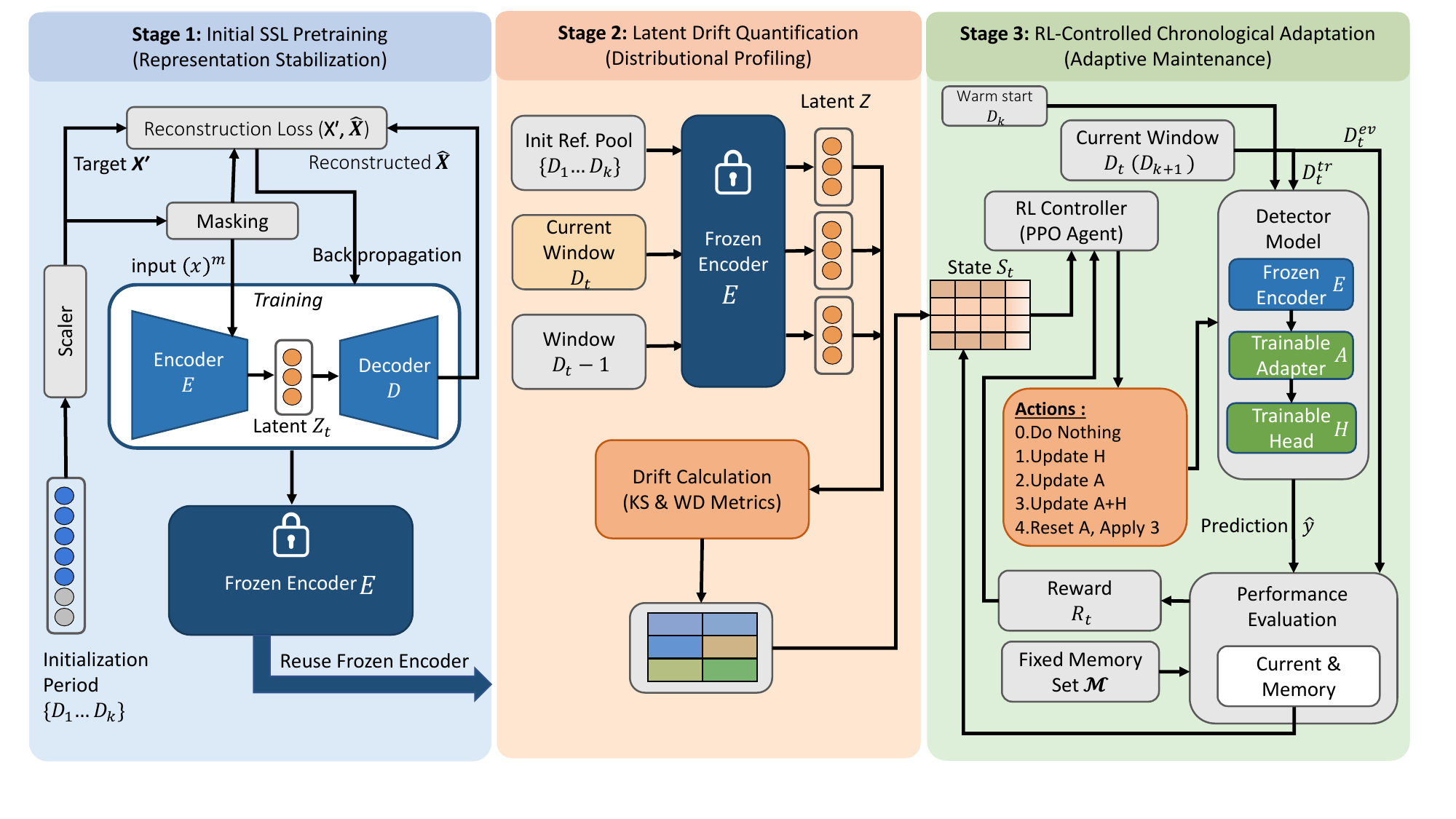}\vspace{-2mm}
    \caption{Proposed stages of maintenance pipeline. Stage~(1) fits the scaler and pretrains the encoder using only the initialization windows $\{\mathcal{D}_1,\ldots,\mathcal{D}_K\}$. Stage~(2) quantifies latent drift in the frozen representation space using previous-window and initialization-reference comparisons. Stage~(3) starts from $\mathcal{D}_{K+1}$ after a warm start on $\mathcal{D}_K$, splits each deployment window into disjoint adaptation and evaluation subsets, and lets a PPO controller choose a maintenance action over the trainable adapter and head while retention is monitored through a fixed memory set $\mathcal{M}$.}
    \label{fig:pipe}\vspace{-2mm}
\end{figure*}

We propose a chronological adaptive maintenance framework for Android malware detection under temporal distribution shift. The framework simulates deployment operation, where a detector is initialized on early time windows and then maintained sequentially as new windows arrive. Instead of repeatedly retraining the full model, the framework learns a stable latent representation once, freezes it, and performs lightweight downstream maintenance using a trainable adapter and classification head controlled by reinforcement learning. As shown in Figure~\ref{fig:pipe}, the framework consists of three stages:
\begin{itemize}
    \item \textbf{Stage 1: Initial SSL pretraining.} A feature scaler and masked self-supervised encoder are trained using the first $K$ chronological windows. After pretraining, the decoder is discarded and the encoder is frozen.
    
    \item \textbf{Stage 2: Latent drift quantification.} For each newly arrived window, drift is measured in the frozen latent space relative to both the previous window and a fixed initialization reference pool.
    
    \item \textbf{Stage 3: RL-controlled chronological maintenance.} Starting from the first post-initialization window, a PPO policy selects among several low-cost maintenance actions that update only the adapter and/or classification head while keeping the encoder frozen.
\end{itemize}

\subsection{Problem definition}

Let $\{\mathcal{D}_t\}_{t=1}^{T}$ denote a chronological sequence of labeled time windows, where
$
\mathcal{D}_t = \{(x_i, y_i)\}_{i=1}^{n_t},
$
$x_i \in \mathbb{R}^{d}$ is the feature vector, and $y_i \in \{0,1\}$ indicates benign or malware class membership. The first $K$ windows, $\{\mathcal{D}_1, \dots, \mathcal{D}_K\}$, define the initialization period used for feature scaling, SSL encoder pretraining, construction of a fixed memory set, and construction of a fixed initialization-reference latent pool. Sequential maintenance begins from $\mathcal{D}_{K+1}$.

At each deployment step, the objective is to maximize current-window detection performance, preserve previously learned behavior, and minimize maintenance cost.

\subsection{Pre-processing and normalization}

After feature selection and chronological partitioning into time windows $\{\mathcal{D}_t\}_{t=1}^{T}$, samples with missing values are removed to preserve a consistent feature space across windows. Each feature vector is normalized using z-score scaling:
$
\tilde{x} = \frac{x - \mu}{\sigma},
$
where $\mu$ and $\sigma$ denote the feature mean and standard deviation. To preserve temporal causality, $\mu$ and $\sigma$ are estimated from the initialization period $\bigcup_{t=1}^{K}\mathcal{D}_t$. The same scaler is then reused unchanged for all subsequent windows.

\subsection{Detector architecture}

The detector consists of three components: $\hat{y} = H\!\left(A\!\left(E(\tilde{x})\right)\right)$, 
where $E(\cdot)$ is a frozen encoder that maps normalized input to a latent representation $z \in \mathbb{R}^{d_z}$, $A(\cdot)$ is a lightweight trainable adapter, and $H(\cdot)$ is a trainable classification head. The encoder is implemented as a multilayer perceptron mapping $d \rightarrow 256 \rightarrow 128 \rightarrow d_z$. The decoder used during SSL pretraining mirrors this architecture in reverse.

The adapter is implemented as a lightweight residual bottleneck transformation:
\begin{equation}
A(z) = z + W_{\mathrm{up}}\!\left(\mathrm{ReLU}\!\left(W_{\mathrm{down}} z\right)\right),
\end{equation}
where \(W_{\mathrm{down}}: \mathbb{R}^{d_z} \rightarrow \mathbb{R}^{b}\) projects the latent representation into a lower-dimensional bottleneck space, \(W_{\mathrm{up}}: \mathbb{R}^{b} \rightarrow \mathbb{R}^{d_z}\) projects it back to the original latent dimension, and \(b \ll d_z\). The residual connection preserves the original representation while allowing the adapter to learn a compact correction.

Freezing the encoder stabilizes the latent space across time, making drift estimates comparable and restricting sequential maintenance to a small number of trainable parameters.

\subsection{Stage 1: Initial SSL pre-training}
\label{subsec:stage1}

The encoder is pretrained during the initialization period using masked feature reconstruction~\cite{LuLFLL25}. For each normalized input \(\tilde{x}\), a binary masking vector \(m \in \{0,1\}^{d}\) is sampled with masking probability \(p\), where \(m_j=1\) indicates that feature \(j\) is masked. The corrupted input is defined as: $\tilde{x}^{(m)} = \tilde{x} \odot (1 - m)$, 
where \(\odot\) denotes element-wise multiplication. The masked input is passed through the encoder to obtain the latent representation: $z = E(\tilde{x}^{(m)})$, 
and the decoder reconstructs the input: $\hat{x} = D(z)$.  The SSL objective is a masked mean-squared reconstruction loss computed only over masked feature positions:
\begin{equation}
\mathcal{L}_{SSL}
=
\frac{\sum_{j=1}^{d} m_j \big(\tilde{x}_j - \hat{x}_j\big)^2}
{\sum_{j=1}^{d} m_j + \epsilon},
\end{equation}
where \(\epsilon > 0\) avoids division by zero.

After pretraining, the decoder is discarded and the encoder is frozen for the remainder of the pipeline. SSL learns stable feature dependencies directly from raw feature vectors without relying on a specific label distribution, reducing the need for repeated representation re-learning under concept drift.

\subsection{Stage 2: Chronological latent drift quantification}
\label{subsec:stage2}

To monitor temporal distribution shift, drift is quantified in the frozen latent space produced by the encoder. For each chronological window \(\mathcal{D}_t\), normalized samples are mapped to latent representations: $Z_t = \{E(\tilde{x}_i)\}_{i=1}^{n_t}$, where \(n_t\) is the number of samples in \(\mathcal{D}_t\). Because the encoder remains fixed after initialization, latent distributions from different windows remain directly comparable over time. At each deployment step, two drift indicators are computed. The first is the previous-window drift:
$
d_{\mathrm{prev}}(t)=\mathrm{Drift}(Z_t, Z_{t-1}),
$
which measures the change relative to the immediately preceding window. The second is the initialization-reference drift:
$
d_{\mathrm{init}}(t)=\mathrm{Drift}(Z_t, Z_{\mathrm{init}}),
$
which measures deviation from a fixed initialization reference pool:
$
Z_{\mathrm{init}} = \bigcup_{i=1}^{K} Z_i.
$

The drift operator combines two complementary distributional distances computed independently for each latent dimension and averaged across dimensions. The first is the mean Kolmogorov-Smirnov (KS) distance:
$$
\mathrm{KS}^{\mathrm{mean}}(Z_a,Z_b)
=
\frac{1}{d_z}
\sum_{k=1}^{d_z}
\sup_u
\left|
F_{a,k}(u)-F_{b,k}(u)
\right|,
$$
where \(F_{a,k}\) and \(F_{b,k}\) are the empirical cumulative distribution functions of the \(k\)-th latent dimension. The second is the mean Wasserstein distance (WD):
$$
\mathrm{WD}^{\mathrm{mean}}(Z_a,Z_b)
=
\frac{1}{d_z}
\sum_{k=1}^{d_z}
W_1\!\left(Z_a^{(k)}, Z_b^{(k)}\right).
$$

The KS distance captures localized distributional differences, while WD captures broader distributional displacement. Using both provides complementary views of latent drift. To keep drift computation tractable and comparable across windows with different sample sizes, drift estimation is performed on a fixed-size subsample of latent vectors from each window.

\subsection{Stage 3: RL-controlled chronological maintenance}
\label{subsec:stage3}

\BfPara{Warm start}
Stage 3 performs chronological maintenance while keeping the encoder frozen. It begins with a supervised warm start on the last initialization window \(\mathcal{D}_K\), where the adapter and classification head are jointly updated. Starting from \(\mathcal{D}_{K+1}\), the RL controller observes the current state, selects a maintenance action, and updates the detector.

\BfPara{Deployment timing}
The framework follows a deployment-style, window-based maintenance setting rather than online sample-by-sample prediction. At step \(t\), a newly labeled window \(\mathcal{D}_t\) becomes available. The controller constructs a chronological state using only information available up to that window, including detector performance, the fixed memory set, and latent drift indicators. A maintenance action is then selected and applied while keeping the encoder frozen. Because the state includes supervised performance on \(\mathcal{D}_t\), the setting represents periodic post-deployment maintenance. 

\BfPara{Disjoint adaptation and evaluation splits}
At each deployment step \(t \geq K+1\), the labeled window \(\mathcal{D}_t\) is partitioned into two disjoint subsets:
$
\mathcal{D}_t = \mathcal{D}_t^{tr} \cup \mathcal{D}_t^{ev},
$
where \(\mathcal{D}_t^{tr}\) is used exclusively for supervised maintenance updates and \(\mathcal{D}_t^{ev}\) is used exclusively for state construction and reward evaluation. This separation prevents optimistic bias from evaluating on samples used for the current update.

\BfPara{Maintenance actions}
At each step, the controller selects one action:
$
a_t \in \{A0,A1,A2,A3,A4\},
$
where \(A0\) performs no update, \(A1\) updates only the classification head, \(A2\) updates only the adapter, \(A3\) jointly updates the adapter and classification head, and \(A4\) reinitializes the adapter before jointly updating the adapter and classification head.

All supervised updates minimize the cross-entropy loss:
\begin{equation}
\mathcal{L}_{CE}
=
-\sum_{c \in \{0,1\}} y_c \log \hat{p}_c,
\end{equation}
where \(y_c\) is the target indicator for class \(c\) and \(\hat{p}_c\) is the predicted probability. If a training budget \(B\) is imposed, updates are performed using at most \(B\) samples from \(\mathcal{D}_t^{tr}\).

\BfPara{Fixed memory set for retention}
To measure retention without storing all previous windows, a fixed class-balanced memory set \(\mathcal{M}\) is sampled from the initialization period and kept unchanged during deployment. Retention at time \(t\) is measured using balanced accuracy on \(\mathcal{M}\), making the retention estimate less sensitive to class imbalance.

\BfPara{State representation}
At deployment step \(t\), the PPO controller observes the following 10-dimensional state vector:
{\small
$$
s_t =
[B_t,F_t,M_t,KS_t^{p},WD_t^{p},KS_t^{i},WD_t^{i},a_{t-1},\mathrm{age}_t,c_{t-1}]
$$
}
where \(B_t\) and \(F_t\) are the pre-update balanced accuracy and macro-F1 on \(\mathcal{D}_t^{ev}\), respectively, and \(M_t\) is the pre-update balanced accuracy on \(\mathcal{M}\). The terms \(KS_t^{p}\) and \(WD_t^{p}\) denote drift relative to the previous window, while \(KS_t^{i}\) and \(WD_t^{i}\) denote drift relative to the initialization reference pool. The remaining variables correspond to the previous action \(a_{t-1}\), the number of deployment windows since the last strong refresh action (\(A3\) or \(A4\)), and the previous action cost \(c_{t-1}\).

\BfPara{Reward function}
After applying action \(a_t\) using \(\mathcal{D}_t^{tr}\), the detector is re-evaluated on \(\mathcal{D}_t^{ev}\) and \(\mathcal{M}\). Let \(B_t^{post}\), \(F_t^{post}\), and \(M_t^{post}\) denote the post-update balanced accuracy, macro-F1, and memory-set balanced accuracy. The reward is:
{\small
$$
R_t=
[
\alpha F_t^{post}
+
(1-\alpha)B_t^{post}
+
\beta M_t^{post}
+
G_t
-
\lambda_c Cost(a_t)
]
$$
}
where \(\alpha\) controls the trade-off between macro-F1 and balanced accuracy, \(\beta\) controls the contribution of memory retention, and \(\lambda_c\) controls the maintenance cost penalty.

The within-step gain term is:
$$
G_t=
\eta_1(B_t^{post}-B_t^{pre})
+
\eta_2(M_t^{post}-M_t^{pre}),
$$
where \(B_t^{pre}\) and \(M_t^{pre}\) are the pre-update balanced accuracy values. In this formulation, drift is not directly penalized; instead, it is provided as part of the state representation.

\BfPara{Policy learning}
The maintenance policy \(\pi_\theta(a \mid s)\) is learned using PPO. At each step, the controller constructs the current state, selects a maintenance action, updates the detector using \(\mathcal{D}_t^{tr}\), evaluates the updated detector on \(\mathcal{D}_t^{ev}\) and \(\mathcal{M}\), computes the reward, and stores the transition for PPO optimization.

\BfPara{Deterministic baselines}
To evaluate the benefit of the RL controller, we compare it against several fixed maintenance strategies. Frozen-init always selects \(A0\), head-tune selects \(A1\), adapter-tune selects \(A2\), and joint-tune selects \(A3\). Periodic-joint(\(k\)) selects \(A3\) every \(k\) deployment windows and \(A0\) otherwise. Drift-Rule selects \(A3\) whenever any drift indicator exceeds a predefined threshold and selects \(A1\) otherwise.

\subsection{Evaluation Protocol}
All policies are evaluated under the same chronological deployment protocol. They use the same initialization period, frozen encoder, feature configuration, fixed memory set, initialization reference pool, adaptation and evaluation split, training budget, and evaluation budget. The proposed PPO controller is compared against fixed maintenance baselines with predefined actions from the same maintenance action set.

Unless otherwise stated, performance is measured after the selected maintenance action has been applied at each deployment step. We use four primary evaluation criteria:
\begin{enumerate}
    \item Current-window balanced accuracy.
    \item Current-window macro-F1.
    \item Memory-set balanced accuracy.
    \item Accumulated maintenance cost.
\end{enumerate}

Balanced accuracy is reported to account for potential class imbalance as it weighs both classes equally through class-wise recall. Macro-F1 is also reported since it captures the class-wise precision-recall tradeoff and is therefore more informative than plain accuracy under possible imbalance.

To summarize temporal performance, we compute the Area Under Time (AUT), following the chronological evaluation protocol used in TESSERACT~\cite{PendleburyPJKC19}. We report AUT for accuracy, macro-F1, and memory-set accuracy. Additionally, the accumulated maintenance cost is computed as $\mathrm{TotalCost}
=
\sum_{t=K+1}^{T}
\mathrm{Cost}(A_t)$, where \(A_t\) denotes the maintenance action selected at deployment step \(t\).

\begin{table}[t]
\caption{Consolidated Experimental Parameters and Settings.}
\label{tab:final_config}\vspace{-1mm}
\centering
\scriptsize
\setlength{\tabcolsep}{3pt}
\renewcommand{\arraystretch}{1.2} 
\begin{tabularx}{\columnwidth}{@{} ll X @{}}
\toprule
\textbf{Category} & \textbf{Item} & \textbf{Setting / Value} \\
\midrule
\textbf{Data} & Setup & 2008--2020; $K=3$ (Init); Seeds: $\{0 \dots 5\}$ \\
\addlinespace[6pt]

\textbf{State} ($s_t$) & Vector ($d=10$) & Curr/Mem Perf, Causal Drift (KS, WD vs. $t-1$ \& Init), Prev. Action, Age, Prev. Cost \\
\addlinespace[6pt]

\textbf{Actions} ($a_t$) & Operators & $A_0$: Keep, $A_1$: Head, $A_2$: Adapter, $A_3$: Joint, $A_4$: Reset+Joint \\
& Update Epochs & Head: 2; Adapter/Joint: 3; Warm start: 4 \\
& Costs ($C$) & $0.0, 0.5, 1.0, 1.5, 2.5$ \\
\addlinespace[6pt]

\textbf{Reward} & Formulation & $R = (\alpha F_1 + (1-\alpha)BA) + \beta BA_{mem} + Gain - \lambda_c C$ \\
& Gain Weights & $\eta_1 (\Delta \text{Curr}) = 1.0, \eta_2 (\Delta \text{Mem}) = 0.5$ \\
& Multipliers & $\alpha, \beta = 0.5$; $\lambda_c = 0.02$ \\
\addlinespace[6pt]

\textbf{PPO} & Architecture & MLP Policy; $n\_steps=64$; Batch=64 \\
& Hyperparams & $lr=3\times10^{-4}$; $\gamma=0.95$; $\text{ent\_coef}=0.01$ \\
\addlinespace[6pt]

\textbf{Buffers} & Budgets & Train/Eval/Mem: 2000 samples (Class Stratified) \\
\bottomrule
\end{tabularx}\vspace{-1mm}
\end{table}

\section{Results and Discussion}\label{sec:Results}
\begin{figure*}[t]
    \centering

    \subfloat[Current-window and memory accuracy (Emulator).\label{fig:emu_acc_mem}]{
        \includegraphics[width=0.48\textwidth]{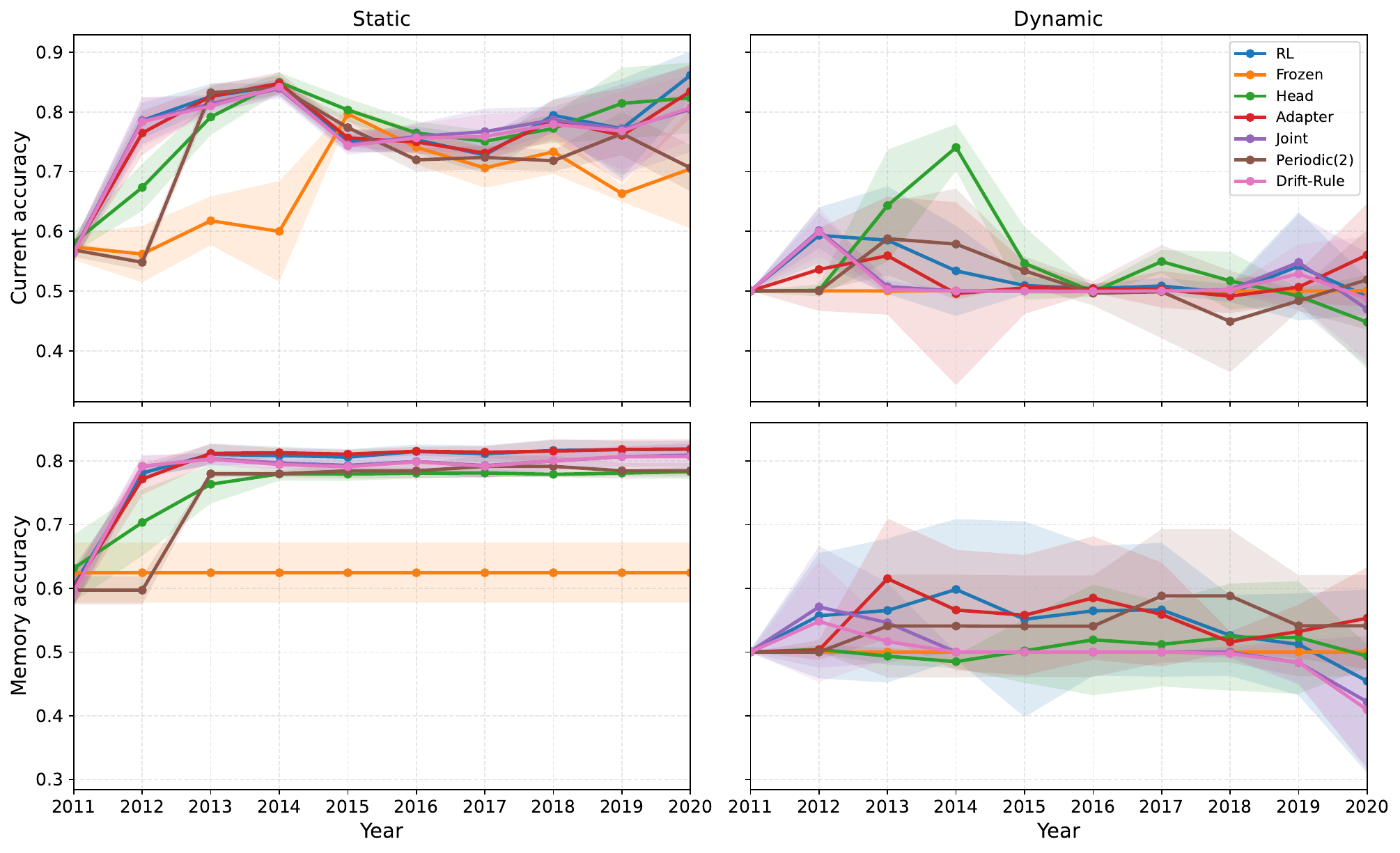}
    }
    \hfill
    \subfloat[Current-window and memory accuracy (Real-data).\label{fig:real_acc_mem}]{
        \includegraphics[width=0.48\textwidth]{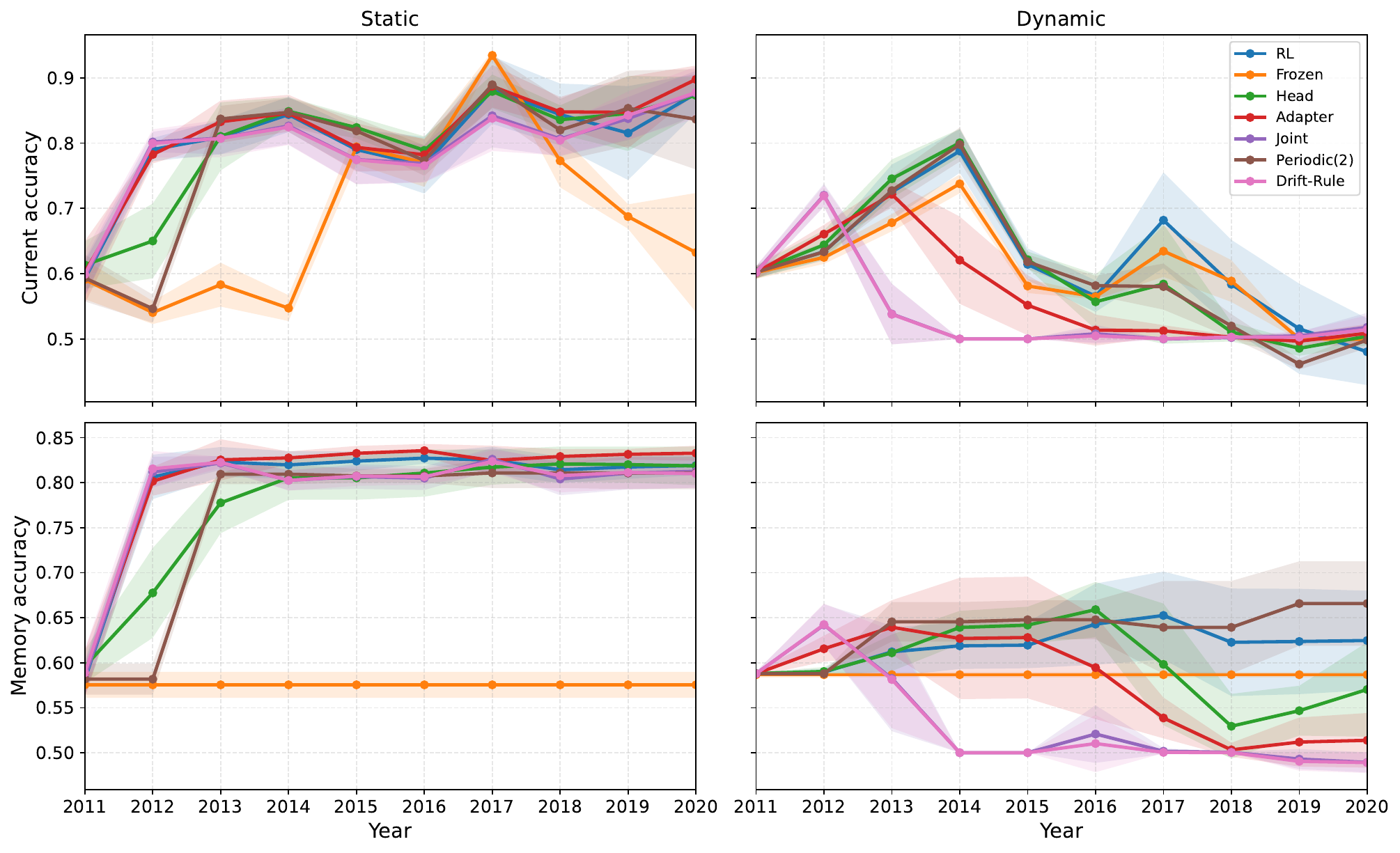}
    }

    \subfloat[Current-window macro-F1 (Emulator).\label{fig:emu_f1}]{
        \includegraphics[width=0.48\textwidth]{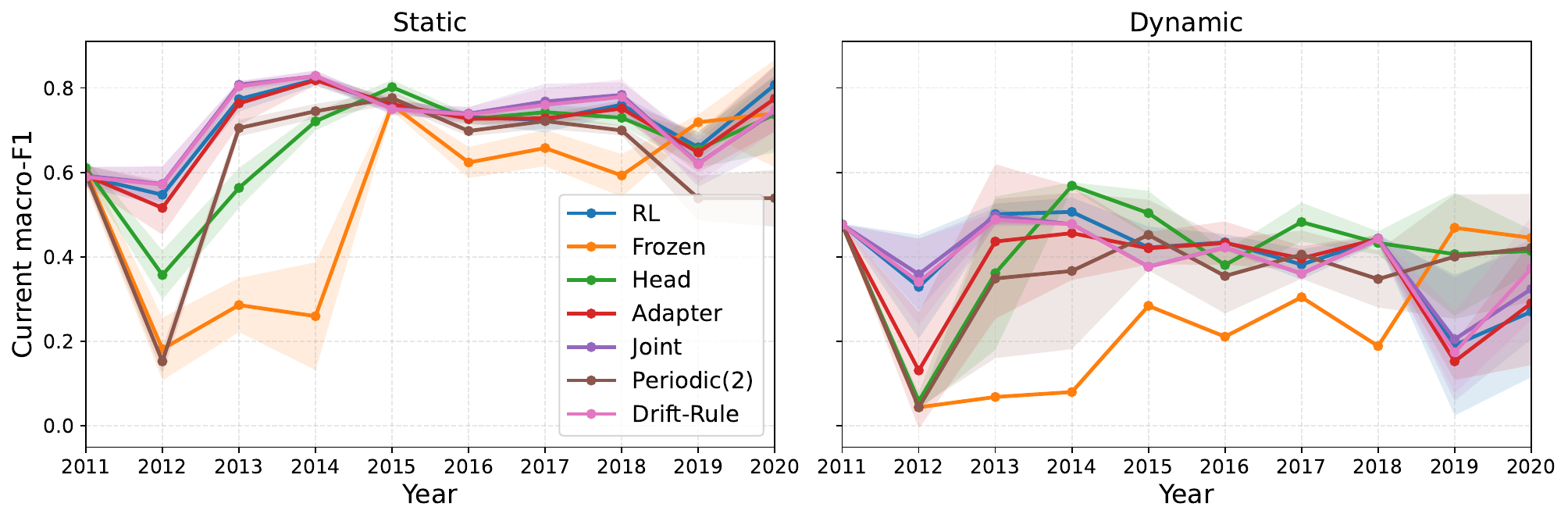}
    }
    \hfill
    \subfloat[Current-window macro-F1 (Real-data).\label{fig:real_f1}]{
        \includegraphics[width=0.48\textwidth]{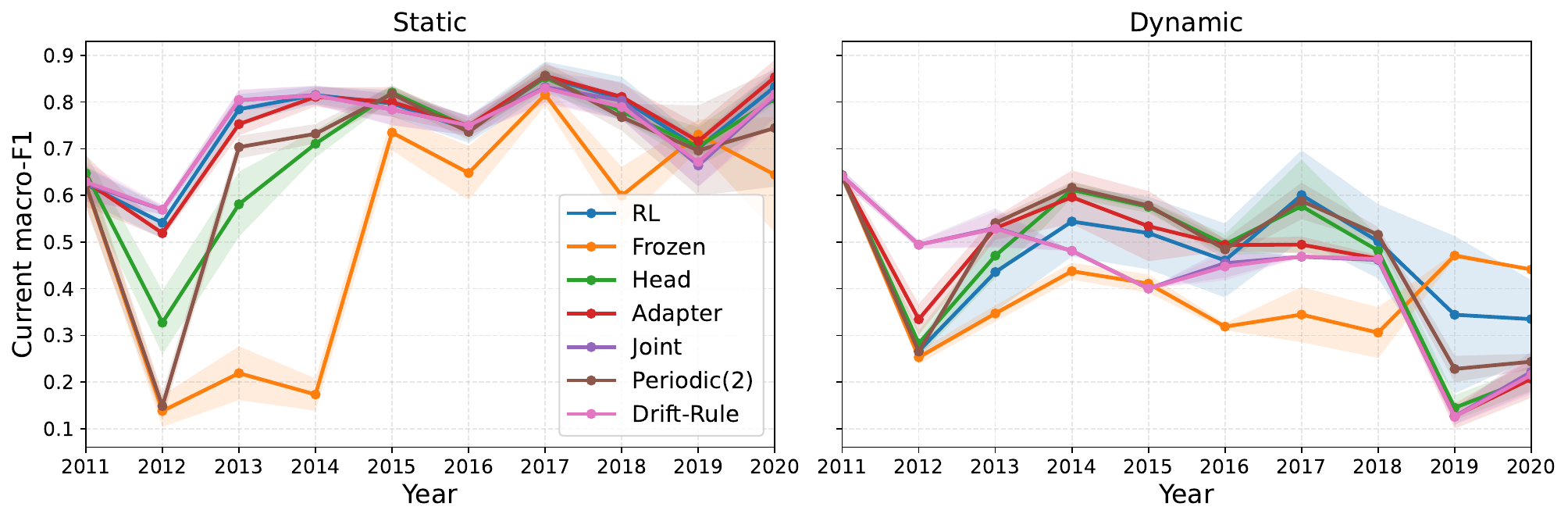}
    }
\vspace{-1mm}
\caption{Temporal performance across emulator and real datasets with static and dynamic features. Accuracy panels report current-window and memory-set performance, while F1 panels report current-window macro-F1.}    \label{fig:all_main_results}\vspace{-1mm}
\end{figure*}

Figure~\ref{fig:all_main_results} shows that temporal performance depends on both the data source and feature type. Static features produce more stable trajectories across emulator and real data, whereas dynamic features yield lower and more variable performance, particularly for current-window macro-F1. Frozen-Init is consistently among the weakest policies, confirming that initialization-only training is insufficient under drift. In contrast, adaptive maintenance policies improve deployment stability. The proposed RL controller is particularly effective because it selects maintenance actions according to the current deployment state, drift indicators, memory behavior, and update cost rather than following a fixed update schedule.

\textit{\BfPara{Takeaway}{
Maintenance during deployment is necessary to address temporal drift. Static features are easier to maintain than dynamic features and the RL controller offers a cost-aware alternative to fixed maintenance schedules by adapting its update decisions to observed deployment states.
}}

A comparison of the adaptive methods shows that no deterministic rule is uniformly optimal across all settings. Head-Tune performs strongly in emulator-dynamic settings, whereas Joint-Tune, Adapter-Tune, and Periodic-Joint(2) remain competitive across several static windows. This is expected because deterministic baselines follow fixed update rules that perform well only when they match the local maintenance needs. In contrast, the proposed RL controller remains competitive across all four settings by selecting maintenance actions according to the current detector state, drift indicators, memory behavior, and update cost. This distinction is most visible in the more challenging dynamic feature settings. In the real-dynamic setting, RL maintains comparatively strong current-window balanced accuracy while keeping memory accuracy relatively stable, whereas several aggressive deterministic rules show sharper late-stage degradation, particularly in macro-F1. More broadly, RL does not need to dominate every year or metric to be useful. Its contribution is replacing handcrafted maintenance rules with a learned, state-dependent controller operating over the same action space. These results support the central claim of this work: under temporal distribution shift, the maintenance policy itself should be adaptive.

\BfPara{Takeaway}~{\em 
RL does not enforce a single preferred update type but learns when different maintenance actions are appropriate. Deterministic baselines are constrained by the limited of their predefined action-selection rules.
}

Table~\ref{tab:aut_summary} summarizes AUT-based current-window balanced accuracy, macro-F1, and accumulated maintenance cost across all settings. No deterministic baseline is uniformly optimal. In emulator-dynamic, Head-Tune achieves the strongest average performance, with approximately 0.551 accuracy and 0.404 macro-F1. In real-dynamic, RL achieves the highest average accuracy, approximately 0.628, whereas Periodic-Joint(2) achieves the highest macro-F1, approximately 0.473. 

The main advantage of RL is not that it dominates every metric, but that it provides a learned, state-dependent maintenance policy over the same action space. This is most visible in real-dynamic, where RL achieves the highest average accuracy while using a substantially lower cost, approximately 3.75, compared with 7.50 for Periodic-Joint(2), 10.00 for Adapter-Tune, and 15.00 for Joint-Tune and Drift-Rule. In real-static, RL achieves the highest average macro-F1, approximately 0.752, whereas Adapter-Tune achieves slightly higher accuracy at lower cost. Similarly, in emulator-static, RL remains competitive, with approximately 0.774 accuracy and 0.719 macro-F1, while the slightly higher F1 values of Joint-Tune and Drift-Rule require the maximum cost of 15.00.

These results support the central claim of this study: under temporal distribution shift, the maintenance policy itself should be adaptive. Deterministic baselines can perform well when their predefined update rules match local maintenance needs, but they remain constrained by fixed behaviors. In contrast, RL learns when to keep the detector unchanged, apply lightweight updates, or perform stronger adaptation based on the observed deployment state and cost.

Another observation is the consistency of RL across seeds. Although RL is not always the top method, its AUT values remain within relatively narrow ranges. For example, in emulator-static, RL maintains accuracy between 0.769 and 0.777 and macro-F1 between 0.707 and 0.732 across the six seeds. In real-dynamic, RL maintains accuracy between 0.618 and 0.648 while using lower cost than the more aggressive Joint-Tune and Drift-Rule baselines. This suggests that the learned controller is not only competitive on average, but also less sensitive to random seed variation than fixed update rules.

\BfPara{Takeaway}~{
\em The AUT results show that RL should be interpreted as a cost-aware and seed-consistent maintenance controller rather than as a method that must win every isolated metric. Across the evaluation settings, the RL remained competitive while avoiding the rigid behavior of the deterministic rules. This supports the use of learned, state-dependent maintenance under temporal distribution shift.
}

\begin{table}[!]
\centering
\caption{AUT summary across seeds and policies. Policy abbreviations are defined as follows: RL: reinforcement learning controller, Fr: Frozen-Init, Hd: Head-Tune, Ad: Adapter-Tune, Jo: Joint-Tune, P2: Periodic-Joint(2), and Dr: Drift-Rule, Acc: Accuracy, C: Cost. S0--S5 are the seeds. Color-coding: \textcolor{worst}{\bf $\blacksquare$}: worst, \textcolor{second}{\bf $\blacksquare$}: second, and \textcolor{best}{\bf $\blacksquare$}: best.}
\label{tab:aut_summary}\vspace{-1mm}
\scriptsize
\setlength{\tabcolsep}{2.0pt}
\renewcommand{\arraystretch}{0.9}

\resizebox{\columnwidth}{!}{%
\begin{tabular}{lcccccccccccc}
\toprule
 & \multicolumn{6}{c}{Emulator} & \multicolumn{6}{c}{Real} \\
\cmidrule(lr){2-7} \cmidrule(lr){8-13}
Policy & \multicolumn{3}{c}{Static} & \multicolumn{3}{c}{Dynamic} & \multicolumn{3}{c}{Static} & \multicolumn{3}{c}{Dynamic} \\
\cmidrule(lr){2-4} \cmidrule(lr){5-7} \cmidrule(lr){8-10} \cmidrule(lr){11-13}
 & Acc & F1 & C & Acc & F1 & C & Acc & F1 & C & Acc & F1 & C \\
\midrule

\multicolumn{13}{l}{\textbf{S0}} \\

\hspace{0.55em}RL &
\second{0.775} & \second{0.718} & 10.500 &
\second{0.508} & \second{0.363} & 12.000 &
\second{0.811} & \second{0.742} & 8.500 &
\best{0.620} & 0.401 & \best{0.000} \\

\hspace{0.55em}Fr &
\worst{0.686} & \worst{0.546} & \best{0.000} &
0.500 & \worst{0.234} & \best{0.000} &
\worst{0.710} & \worst{0.552} & \best{0.000} &
\best{0.620} & 0.401 & \best{0.000} \\

\hspace{0.55em}Hd &
\best{0.793} & 0.689 & \second{5.000} &
\best{0.564} & \best{0.432} & \second{5.000} &
0.783 & 0.662 & \second{5.000} &
\best{0.634} & \second{0.449} & \second{5.000} \\

\hspace{0.55em}Ad &
0.769 & 0.715 & 10.000 &
\worst{0.436} & 0.298 & 10.000 &
\best{0.812} & \best{0.743} & 10.000 &
0.577 & 0.453 & 10.000 \\

\hspace{0.55em}Jo &
0.777 & \best{0.724} & \worst{15.000} &
\second{0.512} & 0.370 & \worst{15.000} &
0.794 & 0.741 & \worst{15.000} &
0.536 & 0.426 & \worst{15.000} \\

\hspace{0.55em}P2 &
0.727 & 0.616 & 7.500 &
0.505 & 0.396 & 7.500 &
0.796 & 0.689 & 7.500 &
0.619 & \best{0.482} & 7.500 \\

\hspace{0.55em}Dr &
0.776 & \best{0.724} & \worst{15.000} &
\best{0.513} & 0.381 & \worst{15.000} &
0.797 & \best{0.743} & \worst{15.000} &
0.533 & 0.422 & \worst{15.000} \\

\addlinespace[1pt]

\multicolumn{13}{l}{\textbf{S1}} \\

\hspace{0.55em}RL &
0.777 & 0.707 & 9.500 &
0.543 & \best{0.451} & \worst{15.000} &
\best{0.821} & \best{0.760} & 10.000 &
\best{0.618} & 0.450 & 6.500 \\

\hspace{0.55em}Fr &
\worst{0.641} & \worst{0.479} & \best{0.000} &
\worst{0.500} & \worst{0.236} & \best{0.000} &
\worst{0.713} & \worst{0.548} & \best{0.000} &
0.606 & \worst{0.380} & \best{0.000} \\

\hspace{0.55em}Hd &
0.753 & 0.632 & \second{5.000} &
\best{0.570} & 0.441 & \second{5.000} &
0.820 & 0.707 & \second{5.000} &
\best{0.618} & \second{0.461} & \second{5.000} \\

\hspace{0.55em}Ad &
0.756 & 0.664 & 10.000 &
0.542 & 0.419 & 10.000 &
\best{0.821} & 0.758 & 10.000 &
0.555 & 0.435 & 10.000 \\

\hspace{0.55em}Jo &
\best{0.787} & \best{0.738} & \worst{15.000} &
0.541 & \second{0.449} & \worst{15.000} &
0.818 & \best{0.761} & \worst{15.000} &
\worst{0.547} & 0.432 & \worst{15.000} \\

\hspace{0.55em}P2 &
0.742 & 0.626 & 7.500 &
0.538 & 0.429 & 7.500 &
0.799 & 0.687 & 7.500 &
0.612 & \best{0.468} & 7.500 \\

\hspace{0.55em}Dr &
\second{0.781} & \second{0.731} & \worst{15.000} &
0.532 & 0.434 & \worst{15.000} &
0.814 & 0.757 & \worst{15.000} &
0.548 & 0.431 & \worst{15.000} \\

\addlinespace[1pt]

\multicolumn{13}{l}{\textbf{S2}} \\

\hspace{0.55em}RL &
0.769 & 0.707 & 9.500 &
0.516 & 0.391 & 10.500 &
0.807 & 0.755 & \worst{15.000} &
\best{0.648} & \second{0.456} & \second{4.000} \\

\hspace{0.55em}Fr &
\worst{0.639} & \worst{0.478} & \best{0.000} &
0.500 & \worst{0.234} & \best{0.000} &
\worst{0.713} & \worst{0.551} & \best{0.000} &
0.586 & \worst{0.349} & \best{0.000} \\

\hspace{0.55em}Hd &
0.751 & 0.645 & \second{5.000} &
\best{0.548} & \best{0.393} & \second{5.000} &
0.811 & 0.712 & \second{5.000} &
0.595 & 0.423 & 5.000 \\

\hspace{0.55em}Ad &
\best{0.775} & \second{0.716} & 10.000 &
\second{0.523} & 0.369 & 10.000 &
\best{0.830} & \best{0.760} & 10.000 &
0.585 & 0.452 & 10.000 \\

\hspace{0.55em}Jo &
0.761 & 0.709 & \worst{15.000} &
0.507 & 0.386 & \worst{15.000} &
0.807 & 0.754 & \worst{15.000} &
0.540 & 0.431 & \worst{15.000} \\

\hspace{0.55em}P2 &
0.716 & 0.614 & 7.500 &
0.511 & 0.302 & 7.500 &
0.803 & 0.705 & 7.500 &
0.607 & \best{0.466} & 7.500 \\

\hspace{0.55em}Dr &
\second{0.767} & \best{0.718} & \worst{15.000} &
\worst{0.498} & \second{0.370} & \worst{15.000} &
0.809 & \second{0.759} & \worst{15.000} &
0.537 & 0.430 & \worst{15.000} \\

\addlinespace[1pt]

\multicolumn{13}{l}{\textbf{S3}} \\

\hspace{0.55em}RL &
\best{0.777} & 0.732 & 10.000 &
0.533 & 0.371 & 10.500 &
0.819 & 0.752 & 10.000 &
\best{0.619} & \best{0.485} & \second{4.000} \\

\hspace{0.55em}Fr &
\worst{0.682} & \worst{0.541} & \best{0.000} &
0.500 & \worst{0.234} & \best{0.000} &
\worst{0.667} & \worst{0.480} & \best{0.000} &
0.615 & \worst{0.395} & \best{0.000} \\

\hspace{0.55em}Hd &
\best{0.774} & 0.680 & \second{5.000} &
\best{0.540} & \second{0.374} & \second{5.000} &
0.795 & 0.685 & \second{5.000} &
0.595 & 0.437 & 5.000 \\

\hspace{0.55em}Ad &
0.768 & 0.726 & 10.000 &
\best{0.572} & \best{0.381} & 10.000 &
\best{0.820} & \best{0.754} & 10.000 &
0.575 & 0.435 & 10.000 \\

\hspace{0.55em}Jo &
0.768 & \best{0.739} & \worst{15.000} &
0.510 & 0.372 & \worst{15.000} &
0.796 & 0.737 & \worst{15.000} &
0.531 & 0.423 & \worst{15.000} \\

\hspace{0.55em}P2 &
0.747 & 0.645 & 7.500 &
0.514 & 0.296 & 7.500 &
0.801 & 0.695 & 7.500 &
0.612 & \second{0.483} & 7.500 \\

\hspace{0.55em}Dr &
0.760 & \second{0.730} & \worst{15.000} &
0.510 & 0.373 & \worst{15.000} &
0.792 & \second{0.730} & \worst{15.000} &
0.530 & 0.420 & \worst{15.000} \\

\addlinespace[1pt]

\multicolumn{13}{l}{\textbf{S4}} \\

\hspace{0.55em}RL &
\best{0.774} & \best{0.718} & 10.000 &
\best{0.521} & \best{0.411} & 9.000 &
0.790 & \best{0.733} & 12.500 &
0.622 & \second{0.474} & \second{5.000} \\

\hspace{0.55em}Fr &
\worst{0.682} & \worst{0.542} & \best{0.000} &
0.500 & \worst{0.234} & \best{0.000} &
\worst{0.689} & \worst{0.515} & \best{0.000} &
0.603 & \worst{0.375} & \best{0.000} \\

\hspace{0.55em}Hd &
0.768 & 0.665 & \second{5.000} &
\second{0.529} & 0.361 & \second{5.000} &
0.789 & 0.674 & \second{5.000} &
\best{0.628} & \best{0.479} & \second{5.000} \\

\hspace{0.55em}Ad &
0.772 & 0.713 & 10.000 &
0.506 & 0.342 & 10.000 &
\best{0.796} & 0.728 & 10.000 &
0.576 & 0.450 & 10.000 \\

\hspace{0.55em}Jo &
\best{0.774} & \second{0.720} & \worst{15.000} &
0.510 & \second{0.397} & \worst{15.000} &
0.776 & \second{0.738} & \worst{15.000} &
0.533 & 0.424 & \worst{15.000} \\

\hspace{0.55em}P2 &
0.713 & 0.609 & 7.500 &
0.503 & 0.357 & 7.500 &
0.772 & 0.669 & 7.500 &
0.608 & 0.464 & 7.500 \\

\hspace{0.55em}Dr &
0.764 & 0.709 & \worst{15.000} &
0.513 & \second{0.398} & \worst{15.000} &
0.774 & \second{0.737} & \worst{15.000} &
0.534 & 0.424 & \worst{15.000} \\

\addlinespace[1pt]

\multicolumn{13}{l}{\textbf{S5}} \\

\hspace{0.55em}RL &
0.774 & \second{0.732} & 11.500 &
\best{0.558} & 0.400 & 10.000 &
0.802 & \best{0.770} & \worst{16.000} &
\best{0.638} & \best{0.509} & \second{3.000} \\

\hspace{0.55em}Fr &
\worst{0.709} & \worst{0.580} & \best{0.000} &
\worst{0.500} & \worst{0.234} & \best{0.000} &
\worst{0.670} & \worst{0.482} & \best{0.000} &
0.611 & \worst{0.388} & \best{0.000} \\

\hspace{0.55em}Hd &
\second{0.775} & 0.669 & \second{5.000} &
\second{0.556} & \best{0.425} & \second{5.000} &
0.820 & 0.725 & \second{5.000} &
0.598 & 0.461 & 5.000 \\

\hspace{0.55em}Ad &
\best{0.777} & \second{0.732} & 10.000 &
0.508 & 0.358 & 10.000 &
\best{0.834} & 0.762 & 10.000 &
0.556 & 0.441 & 10.000 \\

\hspace{0.55em}Jo &
0.768 & 0.731 & \second{15.000} &
0.516 & 0.388 & \worst{15.000} &
0.811 & 0.762 & \worst{15.000} &
0.535 & 0.430 & \worst{15.000} \\

\hspace{0.55em}P2 &
0.728 & 0.627 & 7.500 &
0.521 & 0.333 & 7.500 &
0.762 & 0.649 & 7.500 &
0.588 & \second{0.477} & 7.500 \\

\hspace{0.55em}Dr &
0.770 & \best{0.733} & \worst{15.000} &
0.518 & \second{0.382} & \worst{15.000} &
0.811 & \second{0.764} & \worst{15.000} &
0.535 & 0.431 & \worst{15.000} \\

\bottomrule
\end{tabular}%
}\vspace{-1mm}
\end{table}


\subsection{Threats to Validity}

Our findings are bounded by several factors: (1) the deployment protocol uses yearly chronological windows from 2008--2020 with initialization on the first \(K=3\) windows, so the results support yearly deployment-time maintenance under this chronology rather than arbitrary temporal granularities or deployment schedules; (2) the study assumes a label-available maintenance setting, where each newly labeled window is split into disjoint adaptation and evaluation subsets for state construction, maintenance, and reward computation, rather than fully online adaptation without immediate labels; (3) the comparison uses a fixed maintenance action space and deterministic baselines, with the RL controller trained for 4096 PPO steps, so policy rankings should be interpreted within this controlled design; and (4) the evaluation focuses on current-window balanced accuracy, current-window macro-F1, memory-set balanced accuracy, accumulated maintenance cost, and AUT, and therefore does not assess properties such as calibration, adversarial robustness, or delayed feedback.

\section{Conclusion}\label{sec:conclusion}

This paper presented a chronological adaptive maintenance framework for Android malware detection under temporal distribution shift. The framework learns a stable latent representation during initialization, freezes the encoder, quantifies latent drift in the fixed representation space, and performs lightweight downstream maintenance through an RL-controlled adapter-head update mechanism. Rather than relying on fixed maintenance rules, the proposed RL controller learns state-dependent maintenance decisions based on drift, detector behavior, memory retention, and update cost.

Experiments on emulator and real datasets showed that static features are easier to maintain than dynamic features, that no-maintenance deployment is insufficient under temporal drift, and that no deterministic maintenance rule is uniformly optimal across settings. In contrast, the RL controller remained competitive while using more adaptive and cost-aware maintenance behavior. Future work includes evaluating the framework under additional datasets, temporal granularities, and weaker-feedback deployment settings, as well as expanding the maintenance action space and evaluation criteria.

\end{document}